\documentclass[twocolumn,amsmath,amssymb,prb,floatfix]{revtex4}
\usepackage{graphicx}
\usepackage{dcolumn}
\usepackage{bm}

\begin{document}
\title{Conductance of telescoped double-walled nanotubes from perturbation calculations}
\author{Ryo Tamura}
\affiliation{Faculty of Engineering, Shizuoka University, 3-5-1 Johoku,  Hamamatsu 432-8561, Japan}

\begin{abstract}
In a telescoped double-walled nanotube (TDWNT), with the inner tube partially extracted from the outer tube, the total current is forced to flow between the layers.
Considering the interlayer Hamiltonian as a perturbation, we can obtain an analytic formula for the interlayer conductance.
The accuracy of the perturbation formula is systematically improved by including higher-order terms.
The interlayer interaction effective in the perturbation formula is the product of the interlayer Hamiltonian and the wave function.
It clarifies the effects of 
the spatial range of the interlayer Hamiltonian and the 
 band energy shift.

\end{abstract}
\maketitle



\section{Introduction}
Electronic and mechanical properties of single-walled nanotubes 
(SWNTs) originate from the $\pi$ and $\sigma$ bonds in their honeycomb lattice.\cite{tubeHamada,NT-book,review}
 Although the interlayer bonds are considerably weaker than the intralayer $\pi$ and $\sigma$  bonds, they are important for the formation of double-walled nanotubes (DWNT), multi-walled nanotubes, 
 and nanotube bundles. 
The telescoped double-walled nanotube (TDWNT) shown in Fig.    1 was
formed from a DWNT by partially extracting the inner tube.
Since the rigid honeycomb lattice is relatively unaffected by the weak interlayer bonds,
 the basic motions of the TDWNT
 are limited to the interlayer motions caused 
 by relative slide and rotation
 between the outer and inner SWNTs.
 Attaching a piezoelectric electrode to each edge, where only a single monolayer exists, we can measure 
the relation between  the interlayer motions 
and the $total$  current
forced to flow between
 the layers.
 The interlayer motion and the interlayer force were investigated theoretically using molecular dynamics 
\cite{MD-phonon-sliding-shock-wave,MD-telescope-friction,LDA-interlayer-elastic,MD-GHz-oscillator,DWNT-interlayer-pot-R.Saito}
 and experimentally by AFM and TEM.\cite{AFM-sidecontact-SWNTbridge,AFM-SEM-nano-resonator,AFM-telescope}
The interlayer conductances were
 also measured experimentally.\cite{exp-telescope-Zettl,exp-telescope-Nakayama-conductance}
The relationship between the interlayer conductance
 and the interlayer motion can be used
 to construct nano electro-mechanical systems (NEMS), such as
 nano-mechanical switches, \cite{nano-switch} 
 and nano-displacement sensors.

The interlayer bond between atoms $\alpha$ and $\beta$ 
 is represented by the interlayer Hamiltonian element 
 $H_{\alpha,\beta}$. The interlayer motion influences 
 the conductance through the change of these elements.
 Considering the interlayer Hamiltonian  as a perturbation,
 we can show that the effective interlayer interaction
 is the product of the interlayer  Hamiltonian
 and the wave function.
Though this effective interlayer interaction \cite{ref1} 
 was discussed in Ref.\cite{Tamura},  
its relation to the conductance was complicated.
In the present paper, the perturbation formula 
 simplifies this relation.
The perturbation formula was not discussed
 in most of the preceding theoretical works about 
the conductances  of  TDWNTs 
\cite{Uryu-telescope,telescope-Kim,telescope-Buia,telescope-comp-phys,Kim,telescope-Hasson,telescope-Grace,Tamura}.
In Refs. \cite{NT-FGR,Turnney-Cooper}, the perturbation
 formula was discussed 
 but limited to the incommensurate interlayer configuration.
 The present paper shows that the perturbation
 formula is effective even for commensurate
 TDWNTs, e.g. the (5,5)-(10,10) TDWNT,  despite 
 their larger interlayer interaction.
 In order to confirm the validity of the
 perturbation theory, the effect of higher order terms
 , that was not discussed in Refs. \cite{NT-FGR,Turnney-Cooper},
  is also examined.

 Approximate analytical formulas 
 in Refs. \cite{telescope-Grace,telescope-Kim}
 include fitting parameters other than the Hamiltonian.
 Though the fitting parameters are useful for precise
 reproduction of the exact results, 
 they produce ambiguity
 about the relation between the conductance
 and the Hamiltonian.
 In the present paper, 
 the perturbation formula is discussed for the interpretation of
 the relation between the Hamiltonian and the conductance.
 Thus the fitting parameters are excluded.
 Since there is no fitting parameter,
 agreement between the perturbation formula
 and the exact results is limited.
 Nevertheless the perturbation formula
 is effective in the interpretation when
 it reproduces qualitatively
 the relation between the Hamiltonian
 and the conductance.

The paper is organized as follows. The tight-binding (TB) 
  model used
 in the present paper is described in Sec. II.
The perturbation formulas are derived in Sec. III.
The results of Ref.\cite{Tamura} are reproduced by the perturbation formulas in Sec. IV A. Corrections of the TB
  Hamiltonian  suggested by Refs. \cite{telescope-comp-phys,Kim}
 are analyzed in Sec. IV B and IV C.
Summary and discussion are shown in Sec. V.

\begin{figure}
\includegraphics[width=0.8\linewidth]{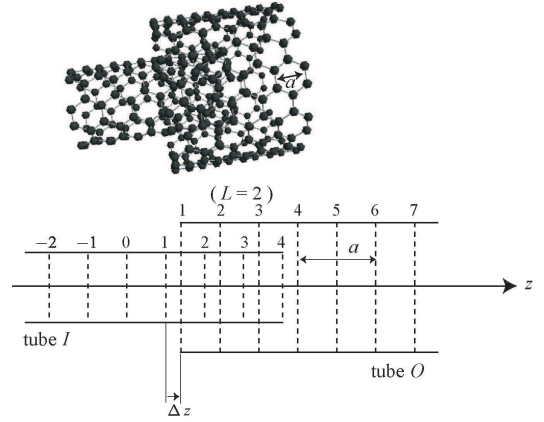}
\caption{ Telescoped double-walled nanotube (TDWNT).
 }
\end{figure}

\section{Tight-Binding Model}
We considered the TDWNT 
composed of two armchair tubes, $(n_O,n_O)$ and $(n_I,n_I)$. 
The symbols $O$ and $I$ 
 indicate the outer and inner tubes, respectively.
As the interlayer distance must be close to that of graphite,
 only the case of $n_O=n_I+5$ was considered. 
Henceforth, the symbol $\mu$ indicates either $O$ or $I$.
The cylindrical coordinates $(r,\theta,z)$ of the atoms 
in the tube $\mu$ are 
\begin{equation}
(r,\theta,z)=
\left(R_{\mu}, \; \frac{\pi \chi_{l,j}}{6n_{\mu}} +\delta_{\mu, I}\Delta \theta,\;\frac{a}{2}j+\delta_{\mu,O}\Delta z\right)
\label{label-O-I}
\end{equation}
with integers $l,j$, $\chi_{l,j} \equiv 6l-5-(-1)^{l+j}$, 
 the lattice constant $a \equiv \sqrt{3} \times 0.142 $  nm,
 and $R_{\mu}=\sqrt{3}an_{\mu}/(2\pi)$.
Regarding the range of $j$,
$j \geq 1$ for tube $O$ 
and $j \leq 2L$ for tube $I$, where $L$ is the number of unit cells 
 in the overlap region.
As shown in Fig.    1,
the overlap length equals $(L-0.5)a -\Delta z$.
 Figure 2 shows the relationship between the integers $(l,j)$
 and the coordinates $(\theta,z)$.
The geometric structure and the definition of
$(\Delta\theta, \Delta z)$
 are the same as in Ref.\cite{Tamura},
where $|\Delta\theta|<\pi/n_O$ and $|\Delta z|<a/4$.

\begin{figure}
\includegraphics[width=0.8\linewidth]{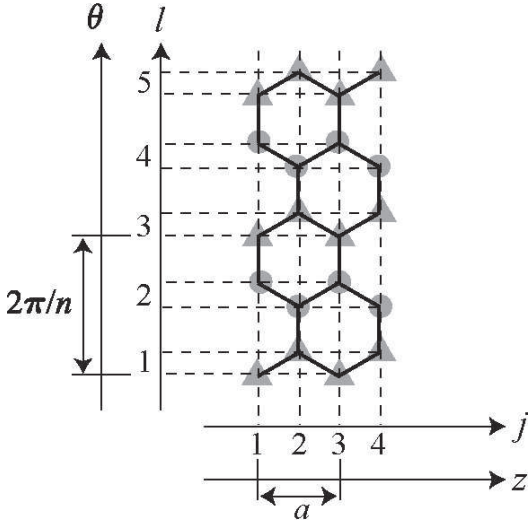}
\caption{
Relationship between integer indexes $(l,j)$ and
 coordinates $(\theta, z)$.
Triangles and circles correspond to odd $l$
 and even $l$, respectively.
 }
\end{figure}

 The $\pi$ orbital at position (\ref{label-O-I})
 is denoted by $|\mu,l,j \,\rangle$
 and is assumed to be 
 orthonormal, i.e., 
 $
\langle \mu^{\;\prime},l^{\;\prime},j^{\;\prime}|
\mu,l,j\,\rangle 
=
 \delta_{\mu,\mu^{\;\prime}} \delta_{l,l^{\;\prime}} \delta_{j,j^{\;\prime}}$, 
where 'bra' and 'ket' are used to simplify the notation.
The Hamiltonian $H$ of the TDWNT is decomposed as
 $H=H^{(0)}_I+H^{(0)}_O+V$, where
$H^{(0)}_{\mu}$ and $V$ 
correspond to
 the intralayer and interlayer Hamiltonian
 elements, respectively.
The interlayer element
between atom $\alpha=(O,l,j)$ and atom $\beta =(I,l^{\;\prime},j^{\;\prime})$ 
 is represented by
\begin{equation}
\langle 
\alpha
|V|
\beta
\rangle 
=\left\{\begin{array}{cc}
(W+\Delta W) e^{\frac{d-r}{L_c}} \cos\theta_{\alpha,\beta}
& \;\cdots r < r_c \\
0 & \;\cdots r > r_c 
\end{array}
\right.
\label{interlayer-H}
\end{equation} 
 with $\theta_{\alpha,\beta}=\theta_{\alpha}-\theta_{\beta}$, inter-atomic distance $r$, constants 
 $W=0.36$ eV, $d=0.334$~nm, $L_c=0.045$~nm, and cutoff radius $r_c$.
 The interlayer bonds were classified as 'AA',
 'BB', or 'AB' bonds.
When either $\alpha$ or $\beta$
had an interlayer bond shorter than
 $r_0$ with the third atom $\gamma$ $( \neq \alpha, \beta)$ 
 and $r >$ $r_0$,
 the bond between $\alpha$ and $\beta$ was classified as
 an AB bond. \cite{Tamura,Lambin,Charlier} 
 Here, $r_0=(a/\sqrt{3})\sqrt{(0.36)^2+(15/2\pi)^2}=
0.34283$ nm,\cite{cut-off-note}
   $\Delta W=-0.20$ eV for AB bonds,
 and $\Delta W=0$ for AA and BB bonds.
 The covalent bond character is the origin of
 the negative value of $\Delta W$, the small cutoff radius $r_c$,
 and the dependence on $\theta$. 
 The intralayer elements were $-t=-2.75$ eV between nearest neighbors,  $\varepsilon$ for the diagonal terms of $H^{(0)}_{O}$, 
 and zero otherwise.
 Even when no interlayer interaction exists,
 the linear dispersion lines of  tube $O$ 
 shift from those of tube $I$
 due to the difference in curvature
 causing  $\sigma - \pi$ mixing.
In the following discussions, this shift is called 'intrinsic'  
and is distinct 
from the shift  induced by interlayer interaction.
\cite{Kim,DWNT-band,single-band-TB-2}
The parameter $\varepsilon$ was introduced
 to represent this intrinsic shift.

The energy  $E^{(\mu)}_{\sigma,k}$ and
the wave function $|\mu,\sigma,k\}$ 
 of an {\it isolated} SWNT
were obtained from   $H^{(0)}_{\mu}|\mu,\sigma,k\}
=E^{(\mu)}_{\sigma,k}|\mu,\sigma,k\}$ 
as 
\begin{equation}
E^{(\mu)}_{\sigma,k}=-2t\cos(ka/2)-t\sigma+\varepsilon\delta_{\mu,O}
\label{SW-E}
\end{equation}
\begin{equation}
|\mu,\sigma,k\}
=\frac{1}{\sqrt{2}}\sum_{j}
\exp(ikaj/2)|\mu,\sigma,j\,\rangle\rangle
\label{SW-wf}
\end{equation}
and
\begin{equation}
|\mu,\sigma,j\,\rangle\rangle
=\frac{1}{\sqrt{2n_{\mu}}}
\sum_{l=1}^{2n_{\mu}}\sigma^{l}
|\mu,l,j\,\rangle 
\label{SW-wf2}
\end{equation}
with the wave number $k$  and the mirror symmetry  $\sigma=\pm 1$.
When  $\varepsilon=0$ and $r_c=(a/\sqrt{3})\sqrt{(1.37)^2+(15/2\pi)^2}=$ 0.39085 nm,\cite{cut-off-note}
 the total Hamiltonian $H=H^{(0)}_I+H^{(0)}_O+V$ becomes equivalent 
 to that of Ref. \cite{Tamura}.
The reflection at the open edges
$(j=1, 2L)$ was neglected here, 
but will be considered in Sec. III B.
The ket $|\mu,\sigma,j\,\rangle\rangle$ defined by Eq. (\ref{SW-wf2}) will be used in Sec. III B.

\section{Derivation of the perturbation formula} 
\subsection{Fermi's golden rule}
Since the interlayer Hamiltonian element in
 Eq.   (\ref{interlayer-H}) is much smaller than 
 the intralayer $\pi$ bonding $t=2.75$ eV,
 it can be considered as a perturbation.
According to Fermi's golden rule (FGR),
the probability of a transition caused by 
 the perturbation $V$ 
per unit time is 
\begin{eqnarray}
P(O,\sigma,k|I,\sigma^{\;\prime},k^{\;\prime})
 & = &\frac{4\pi^2}{h}
\left|
\left\{
O,\sigma,k
|V|
I,\sigma^{\;\prime},k^{\;\prime}
\right\}
\right|^2 \nonumber \\
& & \times \delta(E^{(O)}_{\sigma,k}-E^{(I)}_{\sigma^{\;\prime},k^{\;\prime}}).
\label{P}
\end{eqnarray}
The density of states 
 with {\it positive} group velocity 
 was derived from Eq.   (\ref{SW-E}) 
as
\begin{eqnarray}
D^{(\mu)}_{\sigma}(E^{(\mu)})
& = & \frac{a}{2\pi} \left(\frac{d}{dk}E^{(\mu)}_{\sigma,k}\right)^{-1}
\nonumber \\
& = &
\left(\pi\sqrt{4t^2-(E^{(\mu)}+\sigma t-\varepsilon\delta_{\mu,O})^2}\right)^{-1}
\label{DOS}
\end{eqnarray}
 per {\it  unit cell} of tube $\mu$.
Note that the wave function (\ref{SW-wf})
 was also normalized per {\it unit cell}.

When the Fermi level $E_F$ is close to zero,
 the interlayer current $I$ can be
 estimated to be
\begin{eqnarray}
I & \simeq & 2e\sum_{\sigma,\sigma^{\;\prime}}\int dE^{(O)}
\int dE^{(I)} (f_O-f_I)D^{(O)}_{\sigma}D^{(I)}_{\sigma^{\;\prime}}
\nonumber\\
 & & \times
P(O, \sigma,k|I,\sigma^{\;\prime},k^{\;\prime})
\nonumber\\
 & \simeq
 & 
G_0V_b\sum_{\sigma,\sigma^{\;\prime}}
F_{\sigma,\sigma^{\;\prime}}\left|\left\{O, \sigma,k^{(O)}_{\sigma}|V|I, \sigma^{\;\prime},k^{(I)}_{\sigma^{\;\prime}}
 \right\}\right|^2 ,
\label{I} 
\end{eqnarray}
where
\begin{equation}
F_{\sigma,\sigma^{\;\prime}}\equiv 
4\pi^2 d^{(O)}_{\sigma}d^{(I)}_{\sigma^{\;\prime}},
\end{equation}
$G_0 \equiv 2e^2/h$, $d^{(\mu)}_{\sigma} \equiv D^{(\mu)}_{\sigma}(E_F)$, and $V_b$ denotes the bias voltage.
The Fermi wave number  $k^{(\mu)}_{\sigma}$ 
 satisfies 
$E_F= -2t\cos(k^{(O)}_{\sigma}a/2)-t\sigma+\varepsilon $, $
 E_F= -2t\cos(k^{(I)}_{\sigma^{\;\prime}}a/2)-t\sigma^{\;\prime}$ and the positive group velocity, $\sin(k^{(\mu)}_{\sigma}a/2)>0$.
As both the bias voltage $V_b$ and the temperature
 were close to zero,
 the Fermi-Dirac distribution function difference
 $f_O-f_I$ was replaced by
 $eV_b \delta(E_F-E^{(O)})$
 in Eq.   (\ref{I}).

According to Landauer's formula, on the other hand,
the conductance $G=I/V_b$ is determined by
\begin{equation}
G=G_0\sum_{\sigma=\pm}\sum_{\sigma^{\;\prime}=\pm} 
T_{\sigma,\sigma^{\;\prime}},
\label{Landauer} 
\end{equation}
 where $T_{\sigma,\sigma^{\;\prime}}$
 denotes the interlayer transmission rate
 from $|I,\sigma^{\;\prime},k^{\;\prime}\}$ to $|O,\sigma,k\}$.
By comparing Eq.   (\ref{I}) to Eq.  (\ref{Landauer}), 
an approximate formula for the transmission rate can be obtained
as follows.
\begin{equation}
T_{\sigma,\sigma^{\;\prime}} =
\frac{4}{3t^2}
\left|\left\{O, \sigma, k^{(O)}_{\sigma}
|V|I, \sigma^{\;\prime}, k^{(I)}_{\sigma^{\;\prime}}\right\}\right|^2
\label{perturbation} 
\end{equation}
Here we concentrate our discussion into cases
 where  $|E_F| $ and $|\varepsilon| $
 are much less than $t$, i.e.,
\begin{equation}
k^{(O)}_{\sigma} \simeq k^{(I)}_{\sigma} \simeq 
\frac{\pi}{a}\left( 1+\sigma \frac{1}{3}\right),
\label{k-}
\end{equation}
\begin{equation}
\pi d^{(O)}_{\sigma}
\simeq \pi d^{(I)}_{\sigma^{\;\prime}}\simeq \frac{1}{\sqrt{3}t},
\label{C-new}
\end{equation}
and  $F_{\sigma,\sigma^{\;\prime}} \simeq 4/(3t^2)$.

Using Eqs.  (\ref{SW-wf}),(\ref{SW-wf2}) and (\ref{perturbation}),
\begin{equation}
T_{\sigma,\sigma^{\;\prime}} =
\frac{1}{3t^2}
\left|
A^{(cor)}_{\sigma,\sigma^{\;\prime}}
+
A_{\sigma,\sigma^{\;\prime}}B_{\sigma,\sigma^{\;\prime}}\right|^2 
\label{T}
\end{equation}
 where

\begin{equation}
A^{(cor)}_{\sigma,\sigma^{\;\prime}}\equiv
\left(\frac{\bar{V}_{3,2}^{\sigma,\sigma^{\;\prime}}}
{\omega_{\sigma}^{(O)}}
-
\bar{V}_{2,3}^{\sigma,\sigma^{\;\prime}}
\omega_{\sigma^{\;\prime}}^{(I)}
\right)
\left(
\frac{
\omega_{\sigma^{\;\prime}}^{(I)}}{
\omega_{\sigma}^{(O)}}
\right)^{2L},
\end{equation}

\begin{equation}
A_{\sigma,\sigma^{\;\prime}}\equiv
\sum_{j=1}^{2}
\sum_{s=-1}^{1}
\bar{V}^{\sigma,\sigma^{\;\prime}}_{j,j+s}
\left(\omega_{\sigma^{\;\prime}
}^{(I)}\right)^{j+s}
\left(\omega_{\sigma}^{(O)}\right)^{-j}
\label{A}
\end{equation}

\begin{equation}
B_{\sigma,\sigma^{\;\prime}}\equiv
\sum_{m=0}^{L-1}
\left(
\frac{\omega_{\sigma^{\;\prime}}^{(I)}}
{\omega_{\sigma}^{(O)}}
\right)^{2m},
\label{B}
\end{equation}

\begin{equation}
\omega^{(\mu)}_{\sigma} \equiv \exp(ik^{(\mu)}_{\sigma}a/2),
\end{equation}

and 

\begin{eqnarray}
\bar{V}_{j,j^{\;\prime}}
^{\sigma,\sigma^{\;\prime}}
 & \equiv & \langle\langle O,\sigma,j|V|I,\sigma^{\;\prime},j^{\;\prime} \,\rangle\rangle
\nonumber \\
&=&
\sum_{l=1}^{2n_O}
\sum_{l^{\;\prime}=1}^{2n_I}
\frac{\sigma^{l}\left(\sigma^{\;\prime}\right)^{l^{\;\prime}}
}
{2\sqrt{n_I n_O}}
\langle O,l,j|V|I,l^{\;\prime},j^{\;\prime}\,\rangle .
\label{h(s)}
\end{eqnarray} 
 When $j \leq 0$ or $j^{\;\prime} \geq 2L+1$,
 Eq.  (\ref{h(s)}) equals zero.
 Otherwise Eq.  (\ref{h(s)}) is determined
 by $j-j^{\;\prime}$ and the parity of $j$.
 The cutoff distance $r_c$ in Eq.  (\ref{interlayer-H}) is so short that
 Eq.   (\ref{h(s)}) becomes zero when $|j-j^{\;\prime}| >1$.
In Eq.    (\ref{T}), 
$\{O, k^{(O)}_{\sigma},\sigma|V|I, k^{(I)}_{\sigma^{\;\prime}},\sigma^{\;\prime}\}$ is resolved into 
 the $z$-axis factor $B_{\sigma,\sigma^{\;\prime}}$ 
and $\theta$-axis factor $A_{\sigma,\sigma^{\;\prime}}$
 with the boundary correction $A^{(cor)}_{\sigma,\sigma^{\;\prime}}$
 at $z=(L-0.5)a$.
The correction $A^{(cor)}_{\sigma,\sigma^{\;\prime}}$, however,
 is comparable
 to the $\theta$-factor $A_{\sigma,\sigma^{\;\prime}}$, while 
 the $z$-factor $B_{\sigma,\sigma^{\;\prime}}$ can become much larger
 than unity.
 Thus we can neglect $A^{(cor)}_{\sigma,\sigma^{\;\prime}}$ 
 in Eq.    (\ref{T}) as

\begin{eqnarray}
T_{\sigma,\sigma^{\;\prime}
}
 =  \frac{1}{3t^2}
|A_{\sigma,\sigma^{\;\prime}}|^2|
B_{\sigma,\sigma^{\;\prime}}|^2 .
\label{T-sigma}
\end{eqnarray}
Because $B_{\sigma,-\sigma} \ll B_{\sigma,\sigma}$, 
 $T_{+,-}$ and $T_{-,+}$ 
are negligible compared to $T_{+,+}$ and $T_{-,-}$.
Thus the following discussion will concentrate on 
 the dominant transmission rates, $T_{+,+}$ and $T_{-,-}$.
Since $k^{(O)}a-k^{(I)}a \simeq 2\varepsilon/(\sqrt{3}t)$, 
\begin{equation}
T_{\sigma,\sigma}
 \simeq 
\frac{|A_{\sigma,\sigma}|^2}{\varepsilon^2}
\sin^2\left(\frac{\varepsilon L}{\sqrt{3} t} \right).
\label{conductance2-2}
\end{equation}
When $|\varepsilon|L \ll \sqrt{3}t$ ,
Eq. (\ref{conductance2-2}) is approximated by
\begin{equation}
T_{\sigma,\sigma}
 \simeq \frac{|A_{\sigma,\sigma}|^2}{3t^2}L^2 .
\label{conductance2}
\end{equation}

\subsection{Green's function}
\begin{figure}
\includegraphics[width=0.8\linewidth]{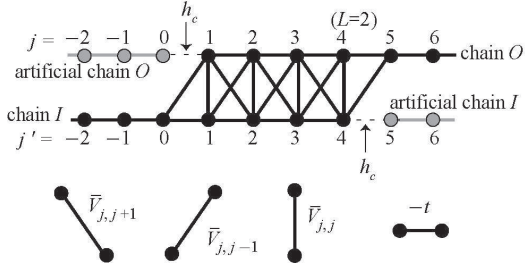}
\caption{\label{ladder}
Schema of the reduced Hamiltonian when $L=2$. }
\end{figure}

With the base set $|\mu ,\sigma,j \,\rangle\rangle $ 
defined by Eq. (\ref{SW-wf2}),
tubes $O$ and $I$ can be approximated by {\it chains}.
The nonzero intra-chain elements are
$\langle \langle  \mu ,j|\,H^{(0)}_{\mu}|\mu,j\pm1\,\rangle\rangle=-t$ 
and
$\langle\langle  \mu ,j|\,H^{(0)}_{\mu}|\mu,j\,\rangle\rangle=
\varepsilon\delta_{\mu,O} -\sigma t$.
Here we suppress index $\sigma$ to simplify
 the notation.
As shown in Fig.    \ref{ladder} ,
$h_c$ was introduced
to cut away the artificial chains and form
 the open edges.
The nonzero elements are
$\langle\langle  I,2L+1\,|\,h_c\,|I^{\,},2L\,\rangle\rangle
=\langle\langle  I,2L\,|\,h_c\,|I^{\,},2L+1\,\rangle\rangle
 =t$
 and
$\langle \langle 
 O,0\,|\,h_c\,|O,1\,\rangle\rangle
  =\langle \langle 
O,1\,|\,h_c\,|O,0\,\rangle\rangle  =t$.
The inter-chain element  $\bar{V}^{\sigma ,\sigma }_{j,j^{\;\prime}}$ was defined by Eq.   (\ref{h(s)}). 
The retarded Green's functions were defined with positive infinitesimal
 $\eta$ as $\widetilde{g}=\left( E+i\eta-H^{(0)}\right) ^{-1}$,
 $g=\left( E+i\eta-H^{(0)}-h_c\right) ^{-1}$,
 $\widetilde{G}=\left( E+i\eta-H^{(0)}-V\right) ^{-1}$,
 and $G=\left( E+i\eta-H^{(0)}-h_c-V\right) ^{-1}$,
 where $H^{(0)}=H^{(0)}_I+H^{(0)}_O$.
The inter-chain elements of $\widetilde{g}$
 and $g$ are zero, while 
 the intra-chain elements $\langle \langle 
\mu ,m\,|\widetilde{g}\,|\mu
,n\,\rangle \rangle 
$ and $ \langle \langle  \mu ,m\,|g\,|\mu
,n\,\rangle\rangle  $ 
are denoted by 
$\left( \widetilde{g}_{\mu}\right) _{m,n}$ and
$\left( g_{\mu}\right) _{m,n}$, respectively.

As was shown in Ref. \cite{Datta},
\begin{equation}
\left( \widetilde{g}_{\mu}\right) _{m,n}
 =\frac{-i}{\sqrt{3}t}
\left( \,\omega
^{(\mu )}\right) ^{|\,m-n\,|}
\label{tilde-g}
\end{equation}
where Eq. (\ref{C-new}) was used.
Using Dyson's equation  $g=\widetilde{g}+\widetilde{g}h_cg$,
we can derive
\begin{equation}
\left( g_{O}\right) _{m,n}
=\left( \widetilde{g}_{O}\right) _{m,n}+\left( 
\widetilde{g}_{O}\right) _{m,0}t\left( g_{O}\right) _{1,n}.
\label{gO-1}
\end{equation}
Substituting $m=1$ in Eq.   (\ref{gO-1}),
\begin{equation}
\left( g_{O}\right) _{1,n}=\frac{\left( \widetilde{g}_{O}\right) _{1,n}}{%
1-t\left(\widetilde{g}_{O}\right) _{1,0}}
\label{gO-2}
\end{equation}
is obtained.
Using Eqs.    (\ref{tilde-g}), (\ref{gO-1}) and (\ref{gO-2}),
\begin{equation}
\left( g_{O}\right) _{m,n}=
\left(\widetilde{g}_O\right)_{m,n}-\left(\widetilde{g}_O\right)_{m+n,0} .
\end{equation}
The matrix $g_I$ is obtained in the same way
as
\begin{equation}
\left( g_{I}\right) _{m,n}=
\left(\widetilde{g}_I\right)_{m,n}-\left(\widetilde{g}_I\right)_{m+n,4L+2} .
\end{equation}
Dyson's equations
 $\widetilde{G}=\widetilde{g}+\widetilde{g}\;V\widetilde{G}=\widetilde{g}\sum_m(V\;\widetilde{g})^m$ 
 and
$G=g+gVG=g\sum_m(Vg)^m$
 indicate
\begin{eqnarray}
\widetilde{G}_{j,j^{\;\prime}}
&= &
\left[ \widetilde{g}_{O}\bar{V}\widetilde{g}_{I}
\left(
 1-^{\;t}\!\bar{V}\widetilde{g}_{O}\bar{V}\widetilde{g}_{I}\right) ^{-1}\right] _{j,j^{\;\prime}}
\end{eqnarray}
and
\begin{equation}
G_{j,j^{\;\prime}}
 =\left[
g_{O}\bar{V}g_{I}\left(
 1-^{\;t}\!\bar{V}g_{O}\bar{V}g_{I}\right) ^{-1}\right]
_{j,j^{\;\prime}}
\end{equation}
,respectively, where
$\widetilde{G}_{j,j^{\;\prime}} \equiv  \langle \langle
 O,j\,|\,\widetilde{G}\,|I^{\,},j^{\;\prime}
\,\rangle \rangle $ 
 and
$G_{j,j^{\;\prime}} \equiv  \langle \langle
 O,j\,|\,G\,|I^{\,},j^{\;\prime}
\,\rangle \rangle $ .
Using Fisher-Lee relation\cite{Datta,Fisher-Lee}
and Eq. (\ref{C-new}),
we can obtain the transmission rate $T$ as
\begin{equation}
T=3t^2\left\vert
\widetilde{G}_{2L+1,0}
 \right\vert ^{2}
\label{T-higher-order}
\end{equation}
including the higher terms
and
\begin{equation}
T=3t^2
\left\vert 
G_{2L+1,0}
  \right\vert ^{2}
\label{T-higher-order-edge}
\end{equation}
 including
 both the higher terms and reflection at the open
 edges.
The explicit relation of Eq.   (\ref{T-higher-order})
 and Eq.   (\ref{T-higher-order-edge})
 to the inter-chain elements $\bar{V}_{j,j^{\;\prime}}$ 
 is summarized in the Appendix.
In the Appendix, we can see that the first order term of 
Eq. (\ref{T-higher-order}),
$3t^2|\left( \widetilde{g}_{O}\bar{V}\,\widetilde{g}_{I} \right)_{2L+1,0}|^2$, coincides with the FGR formula (\ref{T}).

\subsection{Expansion of Eq. (\ref{T-higher-order})}
Equation (\ref{T-higher-order})
 can be expanded as
\begin{equation}
T=3t^2\left|\sum_{m=0}^{\infty}\left(\frac{-1}{3t^2}\right)^{m+1}
q_m\right|^2
\label{T-higher-order2}
\end{equation}
 where
\begin{eqnarray}
q_m &\equiv& (-3t^2)^{m+1}\left( \widetilde{g}_{O}\bar{V}\,\widetilde{g}_{I}(^{\;t}\bar{V}\,\widetilde{g}_{O}\bar{V}\,\widetilde{g}_{I})^{m} \right) _{2L+1,0} \nonumber \\
&=& 
\sum_{j_l,j'_l,s_l,s'_l}
 e^{i(\beta_O+\beta_I)}f
\label{qm}
\end{eqnarray}
\begin{equation}
f = \bar{V}_{j_m,j_m+s_m}\prod_{l=0}^{m-1}\bar{V}_{j_l,j_l+s_l}\bar{V}_{j'_l+s'_l,j'_l}
\label{qm-f}
\end{equation}
\begin{equation}
\beta_O=\frac{k^{(O)}}{2}a\left(-j_0+\sum_{l=0}^{m-1}|j'_l+s'_l-j_{l+1}|\right)
\label{betaO}
\end{equation}
and
\begin{equation}
\beta_I=\frac{k^{(I)}}{2}a\left(|j_m+s_m|+\sum_{l=0}^{m-1}|j_l+s_l-j'_{l}|\right) .
\label{betaI}
\end{equation}
As the ranges of  indexes of Eq. (\ref{qm}) are
$1 \leq j_l \leq 2L,\;1 \leq j'_l \leq 2L,\;  -1 \leq s_l \leq 1$ 
and $ -1 \leq s'_l \leq 1$,\cite{note-index}
the number of terms in Eq. (\ref{qm}) is $3^{2m+1}(2L)^{2m+1}$.
Firstly we consider the case of $\varepsilon=0$,
 i.e., $k^{(O)}=k^{(I)}$.
Among the $3^{2m+1}(2L)^{2m+1}$ terms,  those satisfying
the condition
\begin{equation}
j_0 > j'_0> j_1 >j'_1 > \cdots>j'_{m-1}>j_m
\label{condition}
\end{equation}
 are dominant,
 because
$\beta_O+\beta_I=s_m+\sum_{l=0}^{m-1}(s_l+s'_l)$
 irrespective of    indexes $\{j_l, j'_l\}$ in these dominant terms.
The other terms
cancel each other because of their random phases $\beta_O+\beta_I$.
Since the number of
the dominant terms is $3^{2m+1}(2L)^{2m+1}/(2m+1)!$,
 Eq. (\ref{qm}) can be approximated by
\begin{equation}
q_m=
\frac{L^{2m+1}}{(2m+1)!}A|A|^{2m}
\label{qm2}
\end{equation}
where $A$ was defined by Eq. (\ref{A}).
From Eqs.  (\ref{T-higher-order2}) and (\ref{qm2}),
 we can obtain
\begin{equation}
T=\sin^2\left(\frac{|A|L}{\sqrt{3}t}\right).
\label{T-higher-order3}
\end{equation}

\begin{figure}
\includegraphics[width=0.8\linewidth]{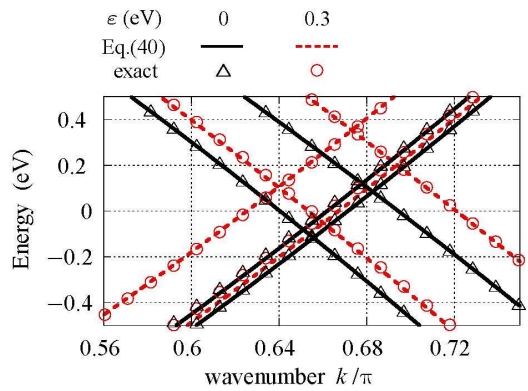}
\caption{\label{band}The dispersion relation of the DWNT with parameters $(n_O, n_I,\Delta\theta,\Delta z)=(10,5,\frac{4\pi}{130}, -0.05a)$  and $\varepsilon$ (eV)  =0, 0.3.
 The triangles and circles show the exact dispersion relation.
 The solid and dashed lines represent Eq. (\ref{disper-doublechain}). }
\end{figure}

In order to discuss the case where  $k^{(O)} \neq k^{(I)}$,
 we refer to the dispersion relation of the DWNT.
 The DWNT can be approximated by the
 double chain  of which the inter-chain Hamiltonian element 
 $V'_{j,j^{\;\prime}}$ is related to Eq. (\ref{h(s)})
 as $V'_{j,j^{\;\prime}}\equiv (\bar{V}_{1,1+j^{\;\prime}-j}+\bar{V}_{2,2+j^{\;\prime}-j})/2$.
 The dispersion relation
 of the double chain is  obtained as \cite{ref1,ref2}
 \begin{equation}
E^{(\pm)}_{\sigma,k}=
E^{(I)}_{\sigma,k}+\frac{1}{2}\left(\varepsilon\pm\sqrt{\varepsilon^2+|\widetilde{A}_{\sigma}|^2}\right)\label{disper-doublechain}
\end{equation}
   where
\begin{equation}
\widetilde{A}_{\sigma}\equiv
\sum_{j=1}^{2}
\sum_{s=-1}^{1}
\bar{V}^{\sigma,\sigma}_{j,j+s}\exp(ikas/2).
\label{Atilde}
\end{equation}
Figure \ref{band} shows that Eq. (\ref{disper-doublechain})  coincides well with the exact dispersion lines.
Comparing Eq. (\ref{Atilde}) with Eq. (\ref{A}),
 we can see that $\widetilde{A}_{\sigma}$ and
 $A_{\sigma,\sigma}$ are essentially the same.
In Eq. (\ref{disper-doublechain}), we can see 
 that the intrinsic band shift $\varepsilon$
 changes  the total band shift $E^{(+)}-E^{(-)}$ 
 from $|A|$ to  $\sqrt{|A|^2+\varepsilon^2}$.
Assuming the same effect of $\varepsilon$ on Eq. (\ref{qm2}),
 we can obtain
\begin{equation}
q_m=
\frac{L^{2m+1}}{(2m+1)!}A(|A|^2+\varepsilon^2)^{m} .
\label{qm3}
\end{equation}
From Eqs.  (\ref{T-higher-order2}) and (\ref{qm3}),
 we can also obtain
\begin{equation}
T=\frac{|A|^2}{|A|^2+\varepsilon^2}
\sin^2\left(\frac{L\sqrt{|A|^2+\varepsilon^2}}{\sqrt{3}t}\right).
\label{T-higher-order4}
\end{equation}
When  $L \sqrt{|A|^2+\varepsilon^2 }  \ll \sqrt{3}t $
 or $|A| \ll |\varepsilon|$, 
Eq. (\ref{T-higher-order4}) coincides with
 Eq. (\ref{conductance2-2}).

\section{Analysis with the perturbation formula }
 In the following sections,
 the transmission rates at $E_F=0.1$ eV $\simeq 0.0364 t$ 
are calculated from
 the conditioned transfer matrix (CTM) \cite{Tamura,Tamura-CTM}
 and the perturbation formulas   (\ref{T-sigma}),(\ref{T-higher-order}),(\ref{T-higher-order-edge}) with the common TB Hamiltonian.
 In the CTM, 
 the transmission rates were obtained from the $S$ (scattering) 
 matrix and the numerical errors were estimated to
 be $\sum_{i,k}|(\sum_j S^*_{j,i}S_{j,k})-\delta_{i,k}|$,
 as the exact $S$ matrix must be unitary.
The estimated errors of the CTM in this paper were less than  $4 \times 10^{-6}$.
Though the perturbation results were less accurate
 than the CTM results, they are useful
 in analyzing the CTM results.

\subsection{Original TB model ( $\varepsilon=0$, $r_c=0.39085$ nm)}

\begin{figure}
\includegraphics[width=0.8\linewidth]{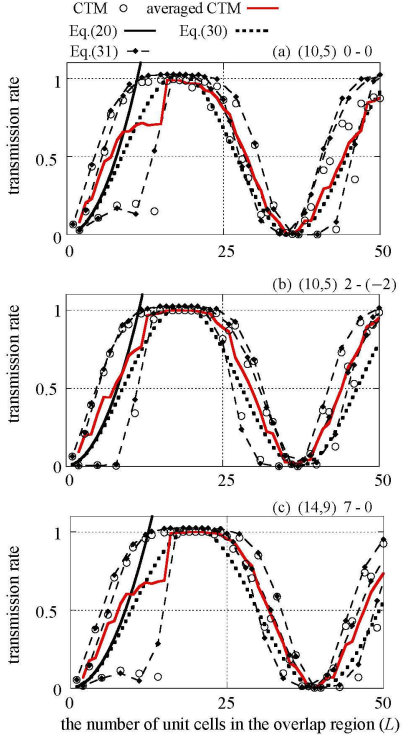} 
\caption{\label{T+} Interlayer transmission rates  $T_{+,+}$ 
 between the $+$ channels
 of $(n_O,n_I)$ $i_{\theta}$-$i_z$ TDWNTs
 at $E_F=0.1$ eV, where
 $\Delta \theta = 2\pi i_{\theta}/(13n_O)$ and
 $\Delta z=a i_z/40$.
 The black solid lines, dotted lines, closed diamonds, 
 open circles and red solid lines  show Eqs.   (\ref{T-sigma}),
 (\ref{T-higher-order}),
 (\ref{T-higher-order-edge}),
 CTM and
 averaged CTM,  respectively.
The closed diamonds are connected with the dashed lines at intervals of $3a$. 
The values of $|A_{+,+}|/t$ are 0.1492, 0.1435 and
 0.1355  in (a), (b) and (c), respectively.}
\end{figure}

\begin{figure}
\includegraphics[width=0.8\linewidth]{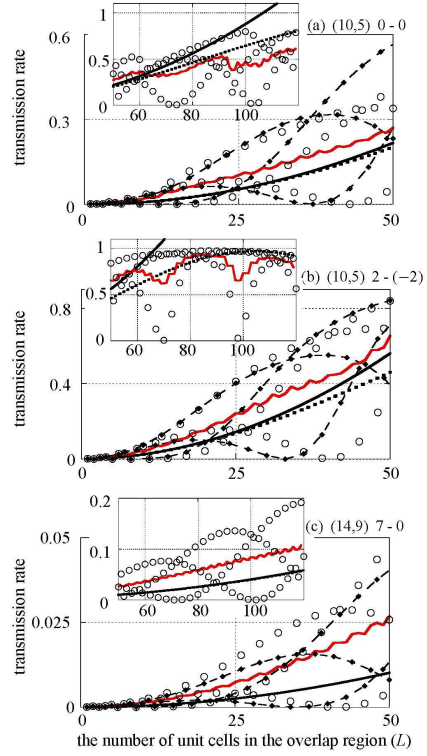} 
\caption{\label{T-} The same calculations as in Fig.    \protect\ref{T+}, except the transmissions  $T_{-,-}$ 
 are between the $-$ channels. 
 In (c), Eq. (\ref{T-sigma}) and Eq. (\ref{T-higher-order})
 nearly coincide with each other.
In the insets with the range 
 $50 \leq L \leq 120$, the data of Eq. (\ref{T-higher-order-edge})
 are omitted because they make the insets busy.
The values of $|A_{-,-}|/t$ are 0.0167, 0.0262 and 0.0036 in (a), (b) and (c),
 respectively.
}
\end{figure}

Figures  \ref{T+} and \ref{T-}
 represent $T_{+,+}$ and $T_{-,-}$, respectively, as a function of
 integer $L$. The overlap length $(L-0.5)a-\Delta z$ was changed discretely and $\Delta z$ was fixed.
 The black solid lines, dotted lines, closed diamonds
 and open circles  show Eqs.   (\ref{T-sigma}),
 (\ref{T-higher-order}),
 (\ref{T-higher-order-edge}),
 and CTM,  respectively.
In contrast to the monotonic increase
 of FGR formula (\ref{T-sigma})  with $L$,
 CTM and Eq. (\ref{T-higher-order-edge})
 showed a rapid oscillation  with a period close to $3a$
  superimposed on a slower oscillation.
Only a  slower oscillation appeared in Eq.  (\ref{T-higher-order})
because Eq. (\ref{T-higher-order})
 does not include the reflection at the open edge.
To show the slower oscillation,
the closed diamonds were connected with the dashed lines at intervals of $3a$
 and the averaged CTM data defined as
 $T^{(ave)}_L=(T_{L-1}+T_{L}+T_{L+1})/3$  were shown by 
 the red solid lines.

The effects of the structure parameters ($L, n_O, \Delta z,\Delta \theta$ ) were reproduced qualitatively by the first order formula
(\ref{T-sigma})  when Eq. (\ref{T-sigma}) is less than unity, i.e., when $L <\sqrt{3}t/|A|$. 
The values of $|A|/t$ are shown
 in Figure captions. 
 Even when $L > \sqrt{3}t/|A|$,
 Eqs.   (\ref{T-higher-order}) 
 and (\ref{T-higher-order-edge}), 
 which include higher order terms, were effective,
 indicating the validity of the
 perturbation formula.
 Note the scale of the vertical axis in Fig.   \ref{T-}.
The transmission rate $T_{-,-}$ of the TDWNTs 
becomes larger  particularly 
 when $n_O=n_I+5=10$. \cite{Tamura}

\subsection{Modification of $r_c$}
\begin{figure}
\includegraphics[width=0.8\linewidth]{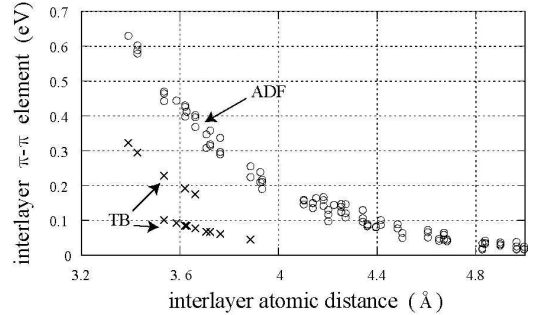} 
\caption{\label{Fock} Interlayer Fock matrix elements between
 $\pi$ orbitals  calculated by ADF.
 Interlayer tight-binding model based on
 Ref.\protect\cite{Lambin} is shown as crosses for comparison. }
\end{figure}

 When $n_O=n_I+5=10$, the maximum conductance
 of the TDWNT is $2G_0$ in the tight-binding (TB) calculation,
 but only $G_0$ in the local density approximation (LDA)
.\cite{Tamura,telescope-Kim,telescope-Buia,telescope-comp-phys,Kim} 
 The difference in the interlayer Hamiltonian
 between TB and LDA is 
 the most probable origin of this disagreement.
To evaluate the order of the interlayer Hamiltonian elements
 in the LDA calculation,
 the ADF calculation \cite{ADF1,ADF2,ADF3} with single zeta 1s,
 2s, and 2p orbitals was performed
 for a (10,10)-(5,5) DWNT composed of four unit cells.
The structure is represented by Eq.    (\ref{label-O-I}), 
 $1 \leq j \leq 8$, 
 $\Delta \theta=0$, and $\Delta z = 0.025a$.
Dangling bonds at $j=1, 8$ were terminated
 by hydrogen atoms
 with a bond length of 0.11 nm.
In the $(R_{\mu}\theta, z)$ plane, the C-C-H angle
 is $2\pi/3$ as is the C-C-C angle.
Geometric optimization was omitted,
 since a slight change in the structure
is not relevant to the order of the interlayer elements.
 With the $\pi$ orbital $\psi_{\pi}$ defined as 
$\psi_{\pi}\equiv \cos\theta \phi_{2{\rm px}}
+\sin\theta \phi_{2{\rm py}}$, 
 Fig.   \ref{Fock} shows the interlayer ADF Fock
 matrix elements of the $\pi$ orbitals
 as a function of
 atomic distance. 
The TB elements used in Ref.\cite{Tamura}
 are also shown in Fig.    \ref{Fock} for comparison.
 The intralayer  elements between nearest neighbors
 were  $-4.73 \sim -4.49$  eV in the ADF  and $-t=-2.75$ eV in the TB.
 Thus, Fig.    \ref{Fock}
  shows that the interlayer elements normalized by the nearest neighbor elements
 were larger   in the ADF than in the TB model.
When we adjust the TB model to reproduce the LDA results,
 the adjusted TB model needs to have contradictory features; 
 {\it larger} interlayer elements and {\it smaller} interlayer
 transmission rates compared to the original TB.
To resolve the contradiction, we should notice that
 the ADF result showed no clear cutoff radius $r_c$
 in Fig.    \ref{Fock}.
 Inspired by these results, we discuss the effect of $r_c$
 in this section.

\begin{figure}
\includegraphics[width=0.8\linewidth]{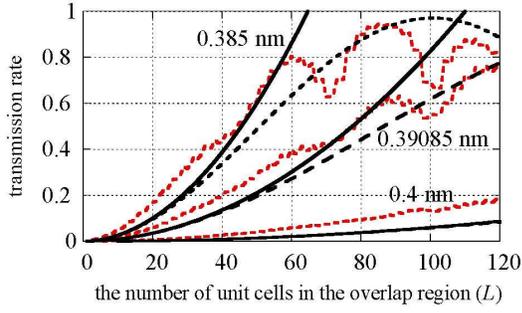}
\caption{\label{rc-T}
Interlayer transmission rate $T_{-,-}$ 
 of the TDWNT ($ n_O=n_I+5=10, \Delta\theta=4\pi/130, \Delta z=0$)
 calculated at $E_F=0.1$ eV using Eq.  (\protect\ref{T-sigma})
 (solid lines),
Eq.   (\ref{T-higher-order}) (dashed lines)
 and the averaged CTM (dotted lines)
 for the shortened cutoff radius $r_c=0.385$ nm,
 the original one $r_c=0.39085$ nm and
  the lengthened one $r_c=0.4$ nm.
 The values of $|A_{-,-}|/t$ are  
 0.0271, 0.0160 and 0.0044
 when $r_c=0.385$ nm,  0.39085 nm and 0.4 nm, respectively. }
\end{figure}

\begin{figure}
\includegraphics[width=0.8\linewidth]{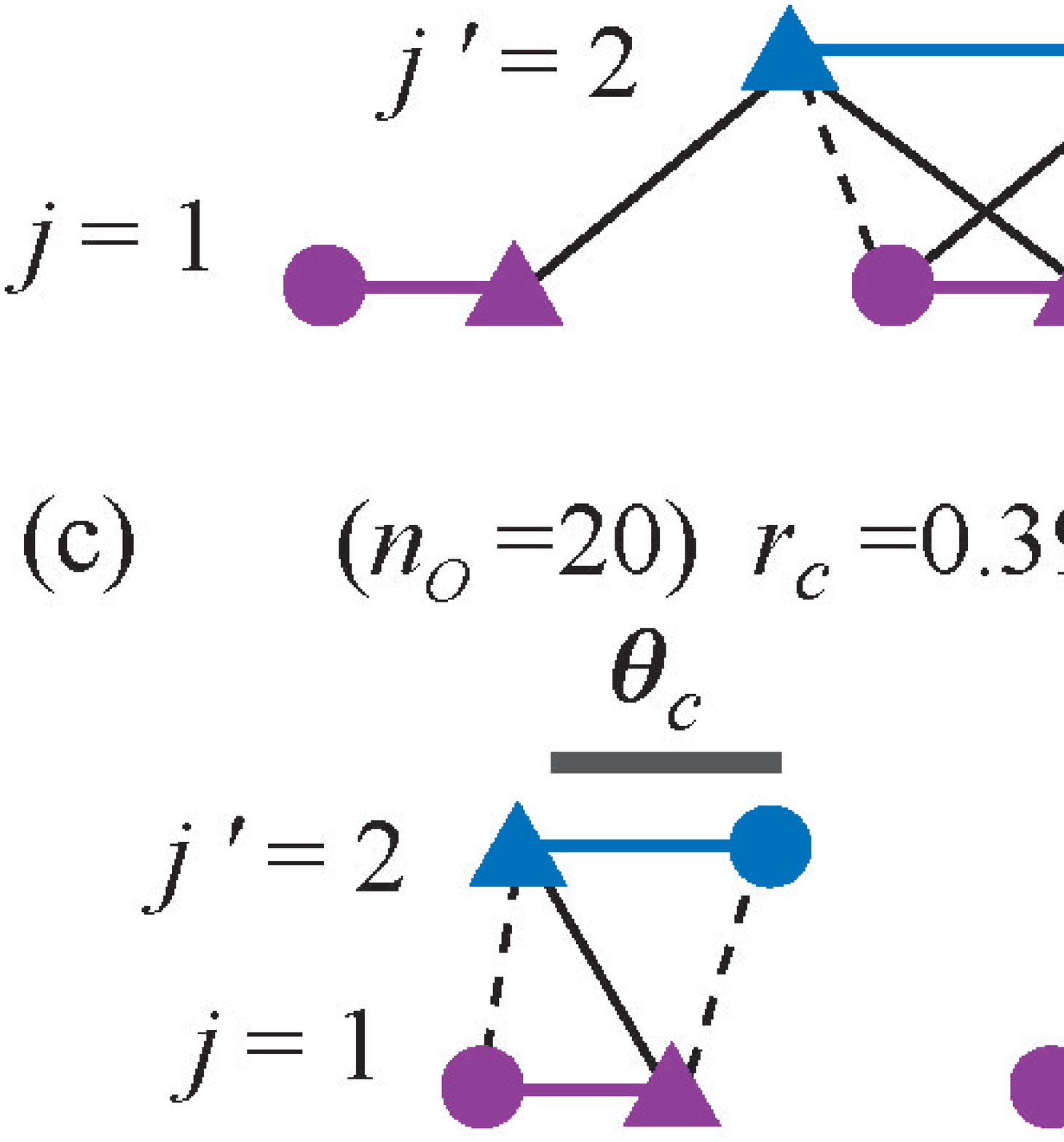} 
\caption{\label{rc-bond}
Horizontal positions of triangles and circles indicate
 $\theta$ coordinates of atoms $(O,l,1)$ and $(I,l^{\;\prime},2)$ 
 (a), (b) $n_O=n_I+5=10, \Delta\theta=4\pi/130, \Delta z=0$, 
 and (c) $n_O=n_I+5=20, \Delta\theta=-\pi/60, \Delta z=0$.
 The interlayer bonds were drawn when the inter-atomic distance was smaller
 than $r_c$, so that the corresponding
 Hamiltonian elements shown by Eq.    (\protect\ref{interlayer-H})
 were finite.
 The cutoff radius $r_c$ was 0.39085 nm in (a) and (c),
 and  0.4 nm in (b).
The interlayer bond between atoms $\alpha$ and $\beta$ 
 was formed 
 only when $|\theta_{\alpha}-\theta_{\beta}|< \theta_c  \equiv 2\arcsin\left(\sqrt{(4\pi^2r_c^2-X))/(12n_In_Oa^2)}\;\right)$.
 Here   $  X \equiv (75+\pi^2(j^{\;\prime}-j)^2)a^2$,  
 $j^{\;\prime}-j=2-1=1$, and $\theta_c$ is shown by bars.}
\end{figure}

 Figure     \ref{rc-T} shows  $T_{-,-}$ for 
 $r_c=0.385$ nm, 0.39085 nm and 0.4 nm.
We can see that $T_{-,-}$ increases
 as $r_c$ decreases
  in the averaged CTM results (dotted lines).
 As this relation between $r_c$ and $T_{-,-}$  was reproduced by
 the  perturbation formulas
  (solid lines and dashed lines),
 it can be explained by 
 the effective
 interlayer interaction $A_{-,-}$ as follows.
 Interlayer bonds are drawn between
 atoms $(O,l,1)$ and $(I,l^{\;\prime},2)$  in Fig.     \ref{rc-bond}
 when the corresponding Hamiltonian elements $ \langle 
O,l,1
|V|
I,l^{\;\prime},2
\,\rangle $ are finite,
 or when the atomic distances are smaller than $r_c$. 
 The interlayer bonds with even or odd $l+l^{\;\prime}$
 are called even or odd bonds in the following discussion,
 and represented by solid or dashed lines, respectively, in Fig.     \ref{rc-bond},
 where the parity of $l$ and $l^{\;\prime}$ is distinguished
 by triangles and circles, using the same representation of parity as in Fig.     2.

\begin{figure}
\includegraphics[width=0.8\linewidth]{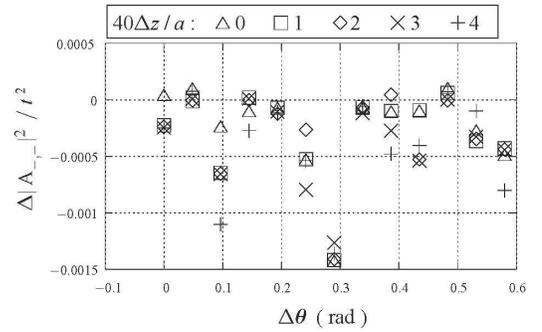}
\caption{\label{Delta-A}
Changes of $|A_{-,-}|^2$
  of the (10,10)-(5,5) TDWNTs 
caused by increase of  $r_c$ from 0.39085 nm
 to 0.4 nm.
The interlayer configurations are 
$\Delta \theta=i_{\theta}\pi/65$
 $(i_{\theta}=0,1,2,\cdots,12)$ and
$\Delta z=i_za/40$
 $(i_z=0,1,2,3,4)$.}
\end{figure}

The parameters 
( $n_O, \Delta\theta, \Delta z$)  were chosen to be the same in Figs.    \ref{rc-bond} (a) and (b)
 as in Fig.     \ref{rc-T}.
 Since $T_{-,-}$ in Eq.    (\ref{T-sigma}) is reduced
 by different parities of $l+l^{\;\prime}$ in Eq.    (\ref{h(s)}), 
 the reduction of $T_{-,-}$ is determined by the balance
 between odd and even bonds.
 This balance tends to be  lost
 when the number of interlayer bonds per unit cell
 is small, as is illustrated by  Fig.     \ref{rc-bond} (a);
 the even bonds are considerably longer than
 the odd bonds,
 although they are equal in number.
 On the contrary, 
 this imbalance is redressed in Fig.     \ref{rc-bond} (b)
 by increase of the number of interlayer bonds caused
 by increasing $r_c$.
 In cases where $n_O > 10$, on the other hand,
  the number of interlayer bonds is large enough
 to reduce $T_{-,-}$  without increasing $r_c$. 
 This is illustrated
 in Fig.     \ref{rc-bond} (c), where $n_O=n_I+5=20,$ $\Delta\theta=-\pi/60$, and $\Delta z=0$.
Figure \ref{Delta-A} shows
$|A_{-,-}(r_c=0.4 {\rm nm})|^2-|A_{-,-}(r_c=0.39085 {\rm nm})|^2$ 
 for the (10,10)-(5,5) TDWNTs of which 
 the interlayer configurations are 
$\Delta \theta=i_{\theta}\pi/65$
 $(i_{\theta}=0,1,2,\cdots,12)$ and
$\Delta z=i_za/40$
 $(i_z=0,1,2,3,4)$.
Figure \ref{Delta-A} clearly indicates
 that increase of $r_c$ tends to reduce $|A_{-,-}|^2$
and $T_{-,-}$.

\subsection{Modification of $\varepsilon$}
Figure 6 of Ref. \cite{Kim} indicates
  a close correlation between
 the intrinsic
 band shift  $|\varepsilon|$ 
 and the suppression of the transmission rate $T_{-,-}$.
These appeared   in multi-band TB \cite{multiband-TB,multiband-TB-okada,Slater}  and LDA,
  but not in  single band TB.
 Here we should distinguish $|\varepsilon|$ from
 the total shift $\sqrt{|A|^2+\varepsilon^2}$ 
 shown by Eq. (\ref{disper-doublechain}).
 Inspired by these results,
 we investigate the effects of $|\varepsilon|$ in this section.

\begin{figure}
\includegraphics[width=0.8\linewidth]{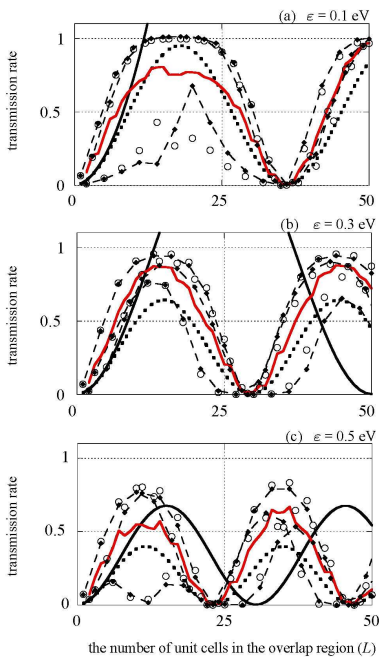}  
\caption{\label{T++E}
The same calculations as in Fig.     \protect\ref{T+} (b), except
 the intrinsic band shift $\varepsilon$ was 
 (a) 0.1 eV, (b) 0.3 eV, or (c) 0.5 eV.
The values of $|A_{+,+}|/t$ are  shown in Table I.
 }
\end{figure}

\begin{figure}
\includegraphics[width=0.8\linewidth]{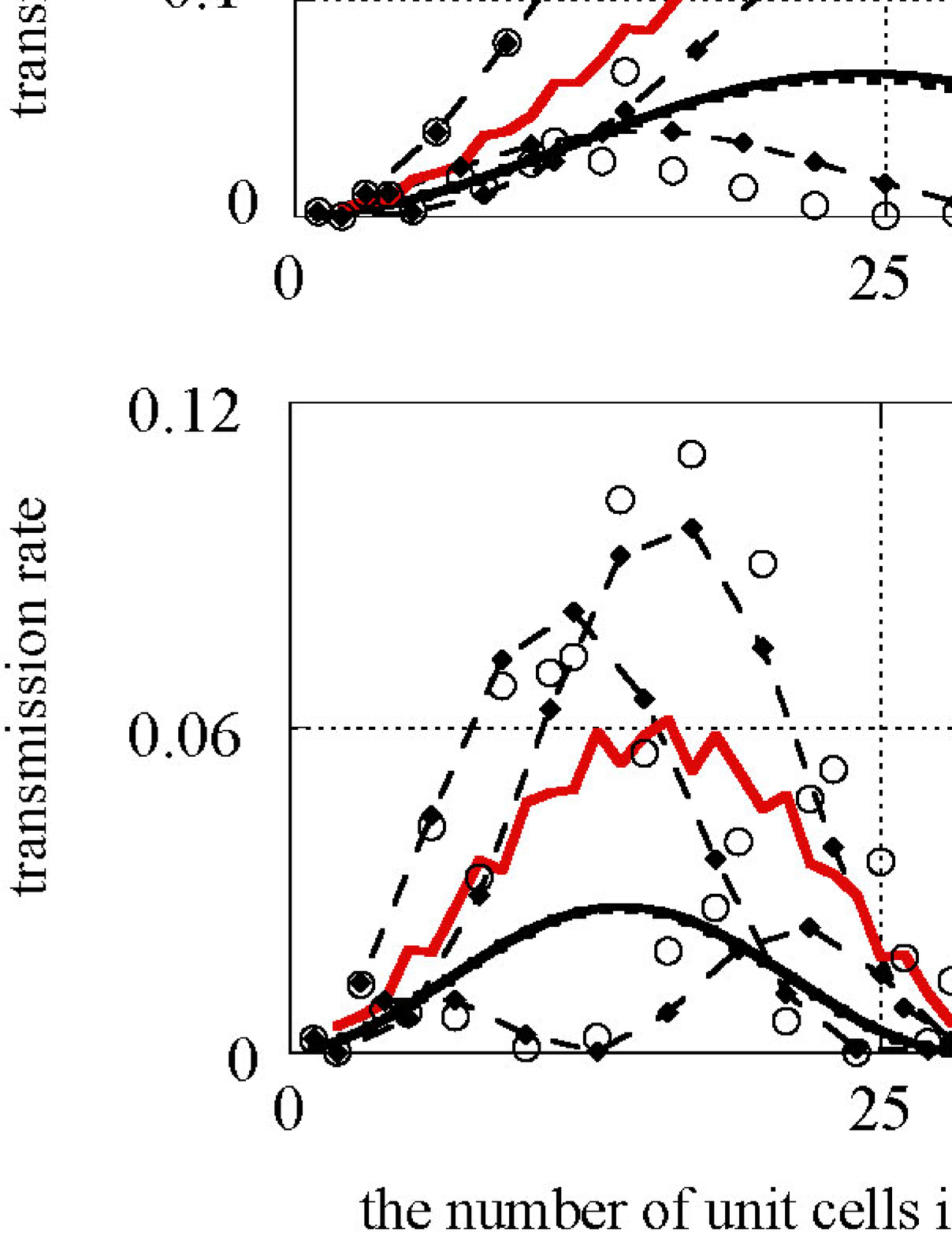} 
\caption{\label{T--E}
The same calculations as in Fig.      \protect\ref{T-} (b), except
 the intrinsic band shift $\varepsilon$ was 
 (a) 0.1 eV, (b) 0.3 eV, or (c) 0.5 eV.
The values of $|A_{-,-}|/t$ are  shown in Table I.
 }
\end{figure}

 Figures 
 \ref{T++E}  and \ref{T--E} show the same 
  calculations as Figs.    \ref{T+} (b)
 and \ref{T-} (b), respectively, except
 that the intrinsic band shift $\varepsilon$ was changed from 
 zero to (a) 0.1 eV (b) 0.3 eV, or (c) 0.5 eV.
 The decrease in the CTM transmission rate
 with increasing $\varepsilon$ 
 was reproduced qualitatively 
 by Eq.    (\ref{T-sigma}).
 We can also see that precision of the  perturbation formula
 was systematically improved by Eqs.   
 (\ref{T-higher-order}) and (\ref{T-higher-order-edge}).

\begin{figure}
\includegraphics[width=0.8\linewidth]{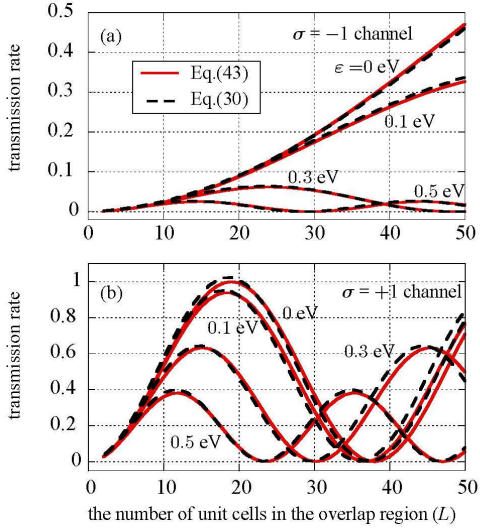}
\caption{\label{expansion}
Comparison between Eq.  (\ref{T-higher-order}) (dashed lines)
 and Eq.  (\ref{T-higher-order4}) (solid lines)
 when 
$(n_O, n_I,\Delta\theta,\Delta z)=(10,5,\frac{4\pi}{130}, -0.05a)$ 
 and $\varepsilon$ (eV)  =0, 0.1, 0.3, 0,5.
We can see that  Eq. (\ref{T-higher-order})
 almost coincides with  Eq. (\ref{T-higher-order4}).
 The data of Eq.  (\ref{T-higher-order})
 are shown also in Figs.  \protect{\ref{T+}}(b),
 \protect{\ref{T-}}(b),
\protect{\ref{T++E}} and \protect{\ref{T--E}}.
}
\end{figure}

 Equation (\ref{T-higher-order})
 almost coincides with Eq. (\ref{T-higher-order4})
 as is seen in Fig.   \ref{expansion}. 
 Thus  Eq.  (\ref{T-higher-order4})
 indicates that
 the first peak of Eq. (\ref{T-higher-order}) as
 a function of $L$ appears at
\begin{equation}
(L,T)=\left(\frac{t\sqrt{3}\pi}{2\sqrt{|A|^2+\varepsilon^2}},
\; \frac{|A|^2}{|A|^2+\varepsilon^2} \right).
\label{firstpeak}
\end{equation}
 The first peak position of the averaged CTM 
 is denoted by $(L_{\rm ctm},\;T_{\rm ctm})$
 and  compared to Eq. (\ref{firstpeak}) in Table I 
 for Figs.  \ref{T+} (b), \ref{T-} (b), \ref{T++E} and \ref{T--E}.
The peak height of Eq. (\ref{firstpeak})
is lower than $T_{\rm ctm}$ when $|\varepsilon|/|A|$ is large.
 Nevertheless Eq. (\ref{firstpeak}) qualitatively reproduces
 the dependence of $(L_{\rm ctm},\;T_{\rm ctm})$
  on $\varepsilon$.
When $|\varepsilon| \ll t $,
 the intrinsic shift $\varepsilon$ exercises only slight influence over $|A|$.


\begin{table}
\begin{ruledtabular}
\begin{tabular}{cccccc} 
\multicolumn{2}{c}{$\varepsilon \;$(eV) }
& 0 & 0.1 & 0.3 & 0.5 \\
\hline

 & & & & & \\

 & $|A_{+,+}|/t$  & 0.1435 & 0.1435 & 0.1432& 0.1430 \\ \cline{2-6}
  & & & & & \\

 & $\frac{\sqrt{3} t \pi}{2\sqrt{|A_{+,+}|^2+\varepsilon^2}}$
  &18.95 &18.38 &15.11  & 11.76 \\

$\sigma=+1$
 & $(L_{\rm ctm})$
  & (18) & (15) & (13) & (14)  \\ \cline{2-6}

 & & & & & \\
 & $\frac{|A_{+,+}|^2}{|A_{+,+}|^2+\varepsilon^2}$
  & 1.000& 0.940 & 0.633 & 0.382  \\ 
 & ($T_{\rm ctm})$
 &  (0.998) & (0.807)  & (0.873) & (0.570)\\
  \hline

 & & & & & \\

 & $|A_{-,-}|/t$  & 0.0262 &  0.0268& 0.0279 & 0.0291  \\ \cline{2-6}
  & & & & & \\

 & $\frac{\sqrt{3} t \pi}{2\sqrt{|A_{-,-}|^2+\varepsilon^2}}$
  &103.9 &60.25 & 24.16  & 14.78 \\ 

$\sigma=-1$
 & $(L_{\rm ctm})$
  & (90) & (70) & (28) & (16)  \\ \cline{2-6}
 & & & & & \\
 & $\frac{|A_{-,-}|^2}{|A_{-,-}|^2+\varepsilon^2}$
 & 1.000 & 0.351 & 0.062 &  0.025  \\ 
 & ($T_{\rm ctm})$
& ( 0.918) & (0.787)  & (0.149) & (0.061)

\end{tabular}
\end{ruledtabular}
\caption{
 The first peak positions 
 of the averaged CTM $(L_{\rm ctm},T_{\rm ctm})$
 in Figs.  \ref{T+} (b), \ref{T-} (b), \ref{T++E} and \ref{T--E}
 are compared with  Eq. (\ref{firstpeak}).
 The values of $|A|/t$ are also shown.}
\end{table}

\section{summary and discussion} 
Considering the interlayer Hamiltonian as a perturbation,
 we derived the  first order formula
(\ref{T-sigma}),
 the formula including the higher-order terms
(\ref{T-higher-order}),
 the formula   including
 both the higher-order terms and reflection at the open
 edges  (\ref{T-higher-order-edge}).
 Expanding Eq. (\ref{T-higher-order}),
 we can see that
 Eq. (\ref{T-higher-order}) is the essentially
 the same as Eq.  (\ref{T-higher-order4}).
They were applied 
 to TDWNTs composed of $(n_O,n_O)$ 
and $(n_O-5,n_O-5)$ armchair tubes. 

 The perturbation formulas clarified
 the effects of the interlayer Hamiltonian
 on the  interlayer conductance $G$.
 The product of the interlayer Hamiltonian
 and  the wave function 
 can be considered as
 the  effective interlayer  interaction
 because it determines
 the perturbation formulas (\ref{T-sigma}), (\ref{T-higher-order4})
 and the dispersion relation (\ref{disper-doublechain}).
 The effective interlayer interaction per unit cell
 and that per total overlapped region
 were denoted by  $A_{\sigma,\sigma^{\;\prime}}$ and
 $A_{\sigma,\sigma^{\;\prime}}B_{\sigma,\sigma^{\;\prime}}$,
 respectively.
Here the parity $\sigma=+,-$ or $\sigma^{\;\prime}=+,-$
 indicates whether the wave function of the outer or inner tube, respectively, changes its sign along the circumference.
 The first order (FGR)  formula (\ref{T-sigma}) was proportional
 to $|A_{\sigma,\sigma^{\;\prime}}|^2|B_{\sigma,\sigma^{\;\prime}}|^2$. Because $|B_{\sigma,-\sigma}|$ 
 was negligible compared to $|B_{\sigma,\sigma}|$,
 we could neglect the inter-channel transmission rates
 $T_{+,-}$ and $T_{-,+}$.

 The CTM transmission rate
 had a rapid oscillation
 superimposed on a slower oscillation
 as a function of  $L$.
 Although Eqs. (\ref{T-sigma})
 and (\ref{T-higher-order})
 could not reproduce the rapid oscillation,
 they approximated the long-period oscillation. 
 The rapid oscillation is
 reproduced by Eq.    (\ref{T-higher-order-edge}),
 which includes
 the reflection at the open edge.
 The first order formula (\ref{T-sigma}) exceeds unity 
 when $|\varepsilon| < |A|$, while the higher-order formula 
 (\ref{T-higher-order})
 never exceeds unity
 and reproduced qualitatively 
 the first peak of the CTM results.
 The systematic improvement of the accuracy indicates the validity of the
 perturbation formulas.

 To represent the range  $L<L_{max}$
 where Eq. (\ref{T-sigma})
 reproduced the averaged CTM results,
 we define $L_1 \equiv t\sqrt{3}/|A|$ and
 $L_2 \equiv \pi t \sqrt{3}/\sqrt{4|A|^2+4\varepsilon^2}$;
 Equation (\ref{conductance2}) reaches unity
 at $L=L_1$ and
 the first peak of Eq. (\ref{T-higher-order})
 appears at $L=L_2$.
 The upper limit of the effective range, $L_{max}$, 
 is classified according to $|\varepsilon|/|A|$ 
as follows.
(i) When $|\varepsilon|/|A| \gg 1$,
Eq. (\ref{T-sigma}) almost coincided
 with Eq. (\ref{T-higher-order})
 and showed the underestimated peak heights.
Nevertheless it reproduced well the period of the oscillation
 of the averaged CTM  even when $L > L_2$.
 In this sense, $L_{max} >  L_2$.
Figures \ref{T--E} (b) and \ref{T--E} (c) correspond to this case.
(ii)$L_{max} = L_2 (< L_1)$ when $|\varepsilon|/|A|$ is
  comparable to unity
 but larger than $\sqrt{\frac{\pi^2}{4}-1} (\simeq 1.2) $.
Figures \ref{T++E} (c) and \ref{T--E} (a) correspond to this case.
(iii)$L_{max} = L_1 (< L_2)$, when $|\varepsilon|/|A|<\sqrt{\frac{\pi^2}{4}-1} $.
Figures \ref{T+},\ref{T-}, \ref{rc-T}, \ref{T++E} (a) and \ref{T++E} (b) correspond to this case.

Since the first principle results suggested the significant effects  of the cutoff radius $r_c$ and the intrinsic band shift $\varepsilon$ on the conductance,
 they were analyzed by the effective interlayer
 interaction $A_{\sigma,\sigma}B_{\sigma,\sigma}$ 
 in Eq. (\ref{T-sigma}).
 The band shift $|\varepsilon|$ reduced the conductance
because it lowered $|B|$.
As $r_c$ became longer,
the number of nonzero terms in Eq.    (\ref{h(s)})
 increased.
 When $\sigma=\sigma^{\;\prime}=-$, 
 the nonzero terms introduced  by increasing $r_c$
 could cancel 
 the terms already present in Eq.    (\ref{h(s)}).
Thus,  a longer $r_c$  could reduce $|A_{-,-}|$ and $T_{-,-}$.

 In contrast to rigorous calculations, which
 involve complicated
 matrix inversion,
 the first order perturbation formula (\ref{T-sigma}) 
 involves
 a simple linear combination of
 the interlayer Hamiltonian elements.
 This enabled us to
 clarify  the role of the interlayer Hamiltonian in 
 the transmission rate.
 In addition to DWNTs,
 there are various other systems composed of two monolayer
 subsystems; side-to-side contact of two SWNTs, \cite{AFM-sidecontact-SWNTbridge,telescope-Buia,Turnney-Cooper}
 an SWNT on graphene,  \cite{NT-on-graphene}
 and bilayer graphene. \cite{few-layer-graphite}
 By sliding one subsystem along the
 other, a generalized telescoped system
 can be obtained in which the interlayer bonds 
 can be considered a perturbation.
 The perturbation formula is an important
 tool to analyze the NEMS  formed
 by these telescoped systems.

\begin{acknowledgements}
The author gratefully acknowledges Prof. Nobuhisa Fujima
 for his advice on ADF.
This work was partly supported by the "True Nano Project" of Shizuoka
University.
\end{acknowledgements}

\appendix*
\section{}
Eqs.  (\ref{T-higher-order})
 and (\ref{T-higher-order-edge})
 are explicitly related to Eq.  (\ref{h(s)}) in this Appendix.
Only the case where $\sigma=\sigma^{\;\prime}$ is considered here
 and thus  index $\sigma$  is suppressed in the following
 formulas as in Sec. III B. For example, the symbol $\bar{V}^{\sigma,\sigma}_{j,j^{\;\prime}}$ is abbreviated as $\bar{V}_{j,j^{\;\prime}}$.

Eqs.  (\ref{T-higher-order}) and (\ref{T-higher-order-edge})
 can be represented by
\begin{equation}
3t^2
\left\vert \widetilde{G}_{2L+1,0}
 \right\vert ^{2}
=\left\vert \sum
\limits_{j^{\,\prime}=0}^{2L}\widetilde{x}_{j^{\,\prime}}\widetilde{Y}_{j^{\,\prime},\,0}\right\vert
^{2}
\end{equation}

and

\begin{equation}
3t^2
\left\vert G_{2L+1,0}
 \right\vert ^{2}
=\left\vert \sum
\limits_{j^{\,\prime}
=0}^{2L}x_{j^{\,\prime}}Y_{j^{\,\prime},\,0}\right\vert ^{2}
\end{equation}

with the following formulas

\begin{eqnarray}
\widetilde{x}_{j^{\,\prime}}
&\equiv&
-\sqrt{3}t
\left(\omega^{(O)}
\right)^{-2L-1}
\left(
\widetilde{g}_O
\bar{V}
\widetilde{g}_I
\right)_{2L+1,j^{\,\prime}}
\nonumber \\
&=&\sum_{j_0=1}^{2L+1}
\left(  \omega^{(O)}\right)
^{-j_0}v_{j_0,j^{\,\prime}}
^{(I)}
\label{appendix-x},
\end{eqnarray}

\begin{eqnarray}
\widetilde{Y}=\left(  1+v^{(O)}
v^{(I)}\right)  ^{-1},
\label{inverse-Y}
\end{eqnarray}

\begin{eqnarray}
v^{(I)}_{j,j^{\,\prime}}
&\equiv&
i(\bar{V}\widetilde{g}_I)_{j,j^{\,\prime}}
\nonumber \\
&=& \sum_{s=-1}^{1}
\frac{\bar{V}_{j,j+s}}{\sqrt{3}t}
\left(\omega^{(I)}\right)^{|j+s-j^{\;\prime}|},
\label{appendix-V}
\end{eqnarray}

\begin{eqnarray}
v^{(O)}_{j^{\,\prime},j}
&\equiv&
i(\,^t\bar{V}\widetilde{g}_O)_{j^{\,\prime},j}
\nonumber \\
&=& \sum_{s^{\;\prime}=-1}^{1}
\frac{\bar{V}_{j^{\;\prime}+s^{\;\prime},j^{\;\prime}}\;}
{\sqrt{3}t}
\left(\omega^{(O)}
\right)^{|j^{\;\prime}+s^{\;\prime}-j|},
\label{appendix-V2}
\end{eqnarray}

\begin{eqnarray}
x_{j^{\,\prime}}
&\equiv&
-i\sqrt{3}t\left(\omega^{(O)}
\right)^{-2L-1}
\left(g_O
\bar{V}g_I
\right)_{2L+1,j^{\,\prime}}
\nonumber \\
&=&\sum_{j=1}^{2L+1}
2\sin\left(\frac{k^{(O)}
a}{2}j\right)X_{j,j^{\,\prime}}^{(I)},
\label{appendix-x2}
\end{eqnarray}

\begin{eqnarray}
Y=(1+X^{(O)}X^{(I)})^{-1},
\label{inverse-Y2}
\end{eqnarray}

\begin{eqnarray}
X_{j,j^{\,\prime}}^{(I)}=v_{j,j^{\,\prime}}^{(I)}
-v_{j,4L+2-j^{\,\prime}}^{(I)},
\end{eqnarray}

and

\begin{eqnarray}
X_{j^{\,\prime},j}^{(O)}=v_{j^{\,\prime},j}^{(O)}-
v_{j^{\,\prime},-j}^{(O)}\;\;.
\end{eqnarray}

For the inverse matrix calculation in Eqs.   (\ref{inverse-Y})
 and (\ref{inverse-Y2}),
$v^{(\mu)}$ and $X^{(\mu)}$ 
are considered as $(2L+1)\times(2L+1)$ 
matrixes in which indexes
 are restricted to $1 \leq j \leq 2L+1$ and
  $0 \leq j^{\;\prime} \leq 2L$.
From Eqs.  (\ref{appendix-x}), (\ref{inverse-Y}),
 (\ref{appendix-V})
 and (\ref{appendix-V2}),
 we can obtain Eqs.  (\ref{T-higher-order2}),
(\ref{qm}), (\ref{qm-f}), (\ref{betaO}) and (\ref{betaI}).
Eqs.  (\ref{appendix-x}) and  (\ref{appendix-V})
 show that 
 the first order term of 
Eq. (\ref{T-higher-order}),
$3t^2|\left( \widetilde{g}_{O}\bar{V}\,\widetilde{g}_{I} \right)_{2L+1,0}|^2=|\widetilde{x}_0|^2$, coincides with the FGR formula (\ref{T}).


\begin{thebibliography}{10}
\bibitem{ref1}
The effective interlayer interaction was denoted by $f(k_{\tau})$ in Ref.\protect\cite{Tamura}.
When $\varepsilon=0$, Eq. (\ref{disper-doublechain}) coincides
 with Eq. (23) of Ref.\protect\cite{Tamura}, where
$k_{\tau}=ka/2,\;t_{\sigma}(\Delta j)
=(\bar{V}^{\sigma,\sigma}_{\Delta j+1,1}
 + \bar{V}^{\sigma,\sigma}_{\Delta j+2,2})/2$
 and $f(k_{\tau})=\widetilde{A}^*_{\sigma}/2$.
\bibitem{cut-off-note}
Five significant figures of $r_0$ and $r_c$ were necessary to reproduce the results of Ref.\protect\cite{Tamura}, because $r_0$ and $r_c$ cause a discontinuous change.
In more realistic calculations, we should replace this by a continuous change.
\bibitem{note-index}
As we neglected $A^{(cor)}$
 in Eq.    (\ref{T}),
 the boundary corrections are neglected
 in  Eq.(\ref{qm}).
In the original definition,
$\bar{V}_{2L+1,2L}=\bar{V}_{1,0}=\bar{V}_{3,2}\neq 0$
and
$\bar{V}_{2L,2L+1}=\bar{V}_{0,1}=0$.
In Eq.(\ref{qm}),
$\bar{V}_{2L+1,2L}=\bar{V}_{1,0}=0$
and
$\bar{V}_{2L,2L+1}=\bar{V}_{0,1}=\bar{V}_{2,3}\neq 0$.
Here the element $\bar{V}_{2,3}$ is the same
 as the original definition.
Figure \ref{ladder} illustrates the original definition
 of $\bar{V}_{j,j^{\;\prime}}$.
\end{thebibliography}
\end{document}